\documentstyle[pra,aps,epsfig]{revtex}
\begin{document}
\draft \twocolumn[\hsize\textwidth\columnwidth\hsize\csname @twocolumnfalse\endcsname
\title{Strong-driving-assisted multipartite entanglement in cavity QED}
\author{E. Solano \footnotemark[1]
$^{1,3}$, G. S. Agarwal$^{1,2}$, and H. Walther$^{1}$}

\address{$^1$Max-Planck-Institut f{\"u}r Quantenoptik,
Hans-Kopfermann-Strasse 1, 85748 Garching, Germany \\ $^2$Physical
Research Laboratory, Navrangpura, Ahmedabad-380 009, India \\
$^{3}$Secci\'{o}n F\'{\i}sica, Departamento de Ciencias,
Pontificia Universidad Cat\'{o}lica del Per\'{u}, Apartado 1761,
Lima, Peru }
\date{\today}
\maketitle

\begin{abstract}
We propose a method of generating multipartite entanglement by
considering the interaction of a system of N two-level atoms in a
cavity of high quality factor with a strong classical driving
field. It is shown that, with a judicious choice of the cavity
detuning and the applied coherent field detuning, vacuum Rabi
coupling produces a large number of important multipartite
entangled states. It is even possible to produce entangled states
involving different cavity modes. Tuning of parameters also
permits us to switch from Jaynes-Cummings to anti-Jaynes-Cummings
like interaction.
\end{abstract}
 \vskip1pc]

\footnotetext[1]{Enrique.Solano@mpq.mpg.de}

Two or more quantum systems are entangled when it is impossible to
describe their physical properties by means of a direct product of
their respective density operators. Entanglement is a natural
consequence of linearity of Hilbert spaces and its controlled
generation and measurement have been intensively
pursued~\cite{Hagley&Turchette}. In a {\it Gedankenexperiment}
after Schr\"odinger~\cite{Schrodinger}, using the properties of
quantum theory, the states of an alive and a dead cat are
correlated with two microscopic states of a decaying atomic
nucleus. The absence of these unusual states in our everyday
experience led to intense debate about the validity of quantum
theory in macroscopic systems. In recent years, great effort has
been put into preparation of the so-called Schr\"odinger cat
states in the laboratory~\cite{HarocheCats,WinelandCats}, where
the extreme cat states have been reduced to mesoscopic quantum
states with classical counterparts, the so-called coherent states.
Decoherence processes, their rate being increased with the size of
the system, have been claimed to be the mechanism inhibiting the
manifestation of entanglement in macroscopic objects. In this
sense, realizing bigger Schr\"odinger cat states in the laboratory
will let us test decoherence by monitoring their
decay~\cite{HarocheCats,WinelandDecoherence}. Meanwhile,
multipartite entangled systems are also considered since they are
interesting in connection with quantum
information~\cite{Sackett,HarocheReview}.

Cavity QED, where atoms interact with a quantized electromagnetic
field inside a cavity, have already proved to be a useful tool for
testing fundamental quantum
properties~\cite{HarocheReview,WaltherReview}. Here, we
demonstrate how a large number of multipartite entangled states
can be generated by adding a strong coherent field to the
system~\cite{Carmichael,Daniel}. The coherent drive affords great
flexibility in generating entangled states since it provides
freedom in choosing the detuning and strength of the field. Our
method enables us to entangle different atoms, cavity modes and
atoms and cavity modes. Furthermore, it is demonstrated how the
external drive has an effect similar to that of a Ramsey field
before and after the cavity~\cite{CqedRamsey}, if we work in
dressed-state basis. It is noteworthy that purely
anti-Jaynes-Cummings like interaction is demonstrated.

The interaction of a two-level atom with a single mode of the
electromagnetic field, described by the Jaynes-Cummings
model~\cite{JaynesCummings}, is one of the simplest and most
fundamental quantum systems. It is typically realized in cavity
QED experiments~\cite{HarocheReview,WaltherReview} in different
frequency regimes and configurations. Here, we consider the
interaction of a single mode (frequency $\omega$) in a cavity of
high quality factor with a spatially narrow bunch of $N$ two-level
atoms (transition frequency $\omega_{o}$), driven additionally by
an external classical field (frequency $\omega_{L}$). The
associated Hamiltonian reads
\begin{eqnarray}
\label{basichamilton} H = \hbar \omega_o \!\! \sum_{j=1}^N \!
{\sigma_{\! j}}^{\dagger} \sigma_{\! j} + \hbar \omega a^{\dagger}
a && \, + \, \hbar \Omega \! \sum_{j=1}^N ( e^{-i \omega_{L} t}
{\sigma_{ \! j}}^{\dagger} \!\! + e^{ i \omega_{L} t} \sigma_{\!
j}) \nonumber \\  && + \hbar g \! \sum_{j=1}^N
({\sigma_{j}}^{\dagger} a + \sigma_{j} a^{\dagger}) ,
\end{eqnarray}
where $\sigma_{j} = | g_{j} \rangle \langle e_{j} |$ and
${\sigma_{j}}^{\dagger} = | e_{j} \rangle \langle g_{j} |$ are the
spin-flip operators down and up, respectively, associated with the
upper level $| e_j \rangle$ and lower level $| g_j \rangle$ of
atom $j$; $a$ and $a^{\dagger}$ are the annihilation and creation
operators associated with the intracavity photon field; $g$ and
$\Omega$, both chosen to be real, are the coupling constants of
the interaction of each atom with the cavity mode and with the
driving field, respectively. A more realistic model of the cavity
mode would include its interaction with a dissipative environment,
finite $Q$, and a thermal bath, finite temperature. Here, we are
interested in the strong coupling regime, $g > \kappa$, where
dissipation can be neglected.

The description of the system of Eq.~(\ref{basichamilton}) is
changed to a reference frame rotating with the driving field
frequency
\begin{eqnarray}
\label{laserframehamilton} H^L = \hbar \Delta \sum_{j=1}^N
{\sigma_{j}}^{\dagger} \sigma_{j} && + \hbar \delta a^{\dagger} a
+ \hbar \Omega \sum_{j=1}^N ( {\sigma_{j}}^{\dagger} + \sigma_{j})
\nonumber
\\ &&  + \hbar g \sum_{j=1}^N
({\sigma_{j}}^{\dagger} a + \sigma_{j} a^{\dagger}) ,
\end{eqnarray}
with $\Delta = \omega_o - \omega_L$ and $\delta = \omega -
\omega_L$. For the sake of simplicity, we set $\Delta = 0$
hereafter. We define
\begin{eqnarray}
&& H^L = H^L_o + H^L_{int} , \nonumber \\ && H^L_o = \hbar \delta
a^{\dagger} a + \hbar \Omega \sum_{j=1}^N ( {\sigma_{j}}^{\dagger}
+ \sigma_{j}) , \nonumber
\\ &&  H^L_{int} = \hbar g \sum_{j=1}^N
({\sigma_{j}}^{\dagger} a + \sigma_{j} a^{\dagger}) .
\end{eqnarray}
The Hamiltonian $H^L$ in the interaction picture yields
\begin{eqnarray}
\label{tildehamilton} H^I = \frac{\hbar g}{2} \sum_{j=1}^N &&
\bigg{ ( } | +_{j} \rangle \langle +_{j} | - | -_{j} \rangle
\langle -_{j} | + e^{2 i \Omega t} | +_{j} \rangle \langle -_{j} |
 \nonumber \\ && -  e^{-2 i \Omega t} | -_{j} \rangle \langle +_{j} | \bigg{ ) } a
e^{- i \delta t} + {\rm H. \, c.},
\end{eqnarray}
where the dressed states $| \pm_{j} \rangle = (| g_{j} \rangle \pm
| e_{j} \rangle) / \sqrt{2}$ are eigenstates of $( \sigma_{x})_j =
\sigma_{j}^{\dagger} + \sigma_{j}$ with eigenvalues $\pm 1$,
respectively. In the strong driving regime $\Omega \gg \{ g ,
\delta \}$, we can realize a rotating-wave approximation and
eliminate from Eq.~(\ref{tildehamilton}) the terms that oscillate
with high frequencies
\begin{eqnarray}
\label{finalhamilton1} H_{e \! f \! f} = && \frac{\hbar g}{2} \!
\sum_{j=1}^N \bigg{ ( } | +_{j} \rangle \langle +_{j} | - | -_{j}
\rangle \langle -_{j} | \bigg{ ) } ( a e^{- i \delta t} +
a^{\dagger} e^{ i \delta t} ) \nonumber \\ = && \frac{\hbar g}{2}
\sum_{j=1}^N (\sigma_{j}^{\dagger} + \sigma_{j}) ( a e^{- i \delta
t} + a^{\dagger} e^{ i \delta t} ).
\end{eqnarray}
A noteworthy feature of Eq.~(\ref{finalhamilton1}), more evident
if $\delta = 0$ and $N = 1$ are chosen, is the simultaneous
realization of Jaynes-Cummings (JC) and anti-Jaynes-Cummings (AJC)
interaction, appearing naturally in trapped ions~\cite{Solano} but
not in the context of cavity QED.

Some examples of the possible applications of the interaction
described in Eq.~(\ref{finalhamilton1}) are given. If at $t=0$,
 $N = 1$, the 1-atom-field state is $| g \rangle | 0 \rangle = (|
+ \rangle + | - \rangle) | 0 \rangle / \sqrt{2}$, the evolved
state after a time $t$ will be
\begin{eqnarray}
\label{cat1} \frac{1}{ \sqrt{2}} ( | + \rangle | \alpha \rangle +
| - \rangle | -\alpha \rangle ) ,
\end{eqnarray}
with $\alpha = g (e^{i \delta t} -1) / 2 \delta$. The
microscopic-mesoscopic entangled state of Eq.~(\ref{cat1}) is
usually called the Schr\"odinger cat state. Clearly, for the
simpler case $\delta = 0$, we have $\alpha = -i g t / 2$, which
shows fast resonant generation of Schr\"odinger cat states as
compared with dispersive methods~\cite{HarocheCats}. Rewriting
Eq.~(\ref{cat1}) in the Schr\"odinger picture,
\begin{eqnarray}
\label{catbare1} \frac{1}{2} \bigg\lbrack && | g \rangle ( e^{-i
\Omega t} | \alpha e^{-i \omega t} \rangle + e^{i \Omega t} |
-\alpha e^{-i \omega t} \rangle) \nonumber \\ && + e^{-i \omega_o
t} | e \rangle ( e^{-i \Omega t}  | \alpha e^{-i \omega t} \rangle
- e^{i \Omega t} | -\alpha e^{-i \omega t} \rangle) \bigg\rbrack ,
\end{eqnarray}
shows that measurement of the atomic state will produce the
so-called even or odd coherent states in the cavity field,
depending on whether $| g \rangle$ or $| e \rangle$ was found,
respectively. Throughout this work, we specify when the final
states are written in the Schr\"odinger picture, as in
Eq.~(\ref{catbare1}), to illustrate the unusual way in which
phases appear after the conveniently realized transformations. If
at $t=0$, $N = 2$, the 2-atom-field state is $| g_1 g_2 \rangle
\otimes | 0 \rangle$, the evolved state after a time $t$ will be
\begin{eqnarray}
\label{cat2} \frac{1}{2} \bigg\lbrack | \phi_1 \rangle | 2 \alpha
\rangle + | \phi_2 \rangle | \! - \! 2 \alpha \rangle + ( | \phi_3
\rangle + | \phi_4 \rangle ) | 0 \rangle \bigg\rbrack ,
\end{eqnarray}
where $| \phi_i \rangle$ are the eigenstates of the atomic
operator $\sum_{j=1}^2 (\sigma_{j}^{\dagger} + \sigma_{j})$ with
eigenvalues $\gamma_{1,2} = \pm 2$ and $\gamma_{3,4} = 0$. The
state of Eq.~(\ref{cat2}) is a bigger and more elaborate
microscopic-mesoscopic 2-atom-field entangled state. Measuring the
two atoms in the state $| g_1 g_2 \rangle$ will produce, in the
Schr\"odinger picture, the field state
\begin{eqnarray}
\label{catbare2}  {\cal N} \bigg( e^{- 2 i \Omega t} | 2 \alpha
e^{-i \omega t} \rangle + e^{2 i \Omega t} | -2 \alpha e^{-i
\omega t} \rangle + 2 | 0 \rangle \bigg) ,
\end{eqnarray}
which is a triple mesoscopic field superposition state with not
only an ``alive" or ``dead" cat, but also and primarily an
``absent" one. The effective interaction of
Eq.~(\ref{finalhamilton1}) lets us create more sophisticated and
bigger microscopic-mesoscopic N-atom-field entangled states and
field superposition states, which will not be described here. Some
of these new states~\cite{Solano}, in fact, present larger quantum
interference regions when displayed in phase-space, showing
distinctly their quantum nature.

When $\delta = \pm 2 \Omega$ and $|\delta| \gg g$,
Eq.~(\ref{tildehamilton}) turns into
\begin{eqnarray}
\label{finalhamilton2} H_{JC}^{(+)} = && \frac{\hbar g}{2}
\sum_{j=1}^N \bigg{ ( } | +_{j} \rangle \langle -_{j} | \, a + |
-_{j} \rangle \langle +_{j} | \, a^{\dagger} \bigg{ ) } ,
\nonumber
\\ H_{AJC}^{(-)} = && \frac{\hbar g}{2}
\sum_{j=1}^N \bigg{ ( } | -_{j} \rangle \langle +_{j} | \, a + |
+_{j} \rangle \langle -_{j} | \, a^{\dagger} \bigg{ ) } ,
\end{eqnarray}
which represent, in the $| \pm_{j} \rangle$ atomic dressed basis,
an effective implementation of a Jaynes-Cummings or an
anti-Jaynes-Cummings interaction. As is known, the anti-JC
interaction does not appear naturally in the context of cavity
QED, where the JC model discards the so-called nonconserving
energy terms, corresponding to exciting (deexciting) the internal
atomic state while creating (annihilating) an intracavity photon.
Another noteworthy feature of the interactions described in
Eqs.~(\ref{finalhamilton2}) is that they produce, in a controlled
manner, absorption and emission of an intracavity photon while
preserving the energy mean value of the atomic state. Clearly, the
apparent imbalance in the energy stems from the external driving
field, which, surprisingly, is intense enough to be considered
classical. Furthermore, let us imagine the situation where the
atom is initially in the ground state $| g \rangle =
\frac{1}{\sqrt{2}}( | + \rangle + | - \rangle )$ and performs a JC
interaction in the atomic dressed basis $| \pm \rangle$, following
any of the Hamiltonians described in Eqs.~(\ref{finalhamilton2}).
Then, by measuring the atom at the end of the interaction in the
atomic bare basis $\{ | g \rangle , | e \rangle \}$, our procedure
is equivalent to a conventional JC evolution with two Ramsey
zones, one before and one after the atom-cavity interaction. This
result makes Ramsey zones unnecessary and may open new
possibilities for phase-sensitive measurements~\cite{Englert} in
closed cavities, as is discussed later.

Let us now concentrate on the possibilities of the studied scheme
in the case of interaction of a bunch of N atoms with two
quasiresonant normal modes in the cavity, always assisted by a
strong external driving field. The associated Hamiltonian, as in
~(\ref{laserframehamilton}), with $\Delta = 0$, reads
\begin{eqnarray}
\label{twomodehamilton} H^{ab} = && \hbar \delta_a a^{\dagger} a +
\hbar \delta_b b^{\dagger} b  + \hbar \Omega \sum_{j=1}^N (
{\sigma_{j}}^{\dagger} + \sigma_{j}) \nonumber
\\ && + \hbar g_a
\sum_{j=1}^N ({\sigma_{j}}^{\dagger} a + \sigma_{j} a^{\dagger}) +
\hbar g_b \sum_{j=1}^N ({\sigma_{j}}^{\dagger} b + \sigma_{j}
b^{\dagger}) ,
\end{eqnarray}
where $\{ a^{\dagger},b^{\dagger} \}$ and $\{ a,b \}$ are the
creation and annihilation operators associated with the two cavity
modes, while  $\delta_a = \omega_a - \omega_L$ and $\delta_b =
\omega_b - \omega_L$. We now define
\begin{eqnarray}
H^{ab} = H^{ab}_o + H^{ab}_{int}
\end{eqnarray}
with
\begin{eqnarray}
\label{interactionpicturetwomode} && H^{ab}_o = \hbar \delta_a
a^{\dagger} a + \hbar \delta_b b^{\dagger} b  + \hbar \Omega
\sum_{j=1}^N ( {\sigma_{j}}^{\dagger} + \sigma_{j}) , \nonumber
\\ && H^{ab}_{int} = \hbar g_a
\sum_{j=1}^N ({\sigma_{j}}^{\dagger} a + \sigma_{j} a^{\dagger}) +
\hbar g_b \sum_{j=1}^N ({\sigma_{j}}^{\dagger} b + \sigma_{j}
b^{\dagger}) .
\end{eqnarray}
In the interaction picture, defined by the formal separation in
Eq.~(\ref{interactionpicturetwomode}), the Hamiltonian changes to
\begin{eqnarray}
\label{tildetwomodehamilton} \tilde{H}^{ab} = \frac{\hbar}{2}
\sum_{j=1}^N && \bigg{ ( } | +_{j} \rangle \langle +_{j} | - |
-_{j} \rangle \langle -_{j} | + e^{2 i \Omega t} | +_{j} \rangle
\langle -_{j} |
 \nonumber \\ && -  e^{-2 i \Omega t} | -_{j} \rangle \langle +_{j} | \bigg{ ) } ( g_a a
e^{- i \delta_a t} + g_b b e^{- i \delta_b t} ) \nonumber \\ && +
\,\, {\rm H. \, c.}
\end{eqnarray}
In the strong driving limit, $\Omega \gg \{ g , \delta_a ,
\delta_b \}$, we obtain
\begin{eqnarray}
\label{twomodefinalhamilton} && H_{ef \! f}^{ab} \!\! = \!\!
\frac{\hbar}{2} \!\! \sum_{j=1}^N (\sigma_x)_{\! j} [g_a ( a e^{-
i \delta_a t} \!\!\! + \! a^{\dagger} e^{ i \delta_a t}  ) \!\! +
\! g_b ( b e^{- i \delta_b t} \!\!\! + \! b^{\dagger} e^{ i
\delta_b t} ) ] . \nonumber \\ &&
\end{eqnarray}
This Hamiltonian will produce states with features similar to
those produced by the one of Eq.~(\ref{finalhamilton1}), only that
now the displacement will act simultaneously on each of the two
cavity modes. If at $t=0$, with $N = 1$ and $\delta_a = \delta_b =
0$, the initial atom-field state is $| g \rangle | 0 \rangle | 0
\rangle = ( | + \rangle + | - \rangle ) | 0 \rangle | 0 \rangle /
\sqrt{2} $, the evolved state at time $t$ will be
\begin{eqnarray}
\label{twomodecat} \frac{1}{ \sqrt{2}} \bigg( | + \rangle | \alpha
\rangle | \beta  \rangle + | - \rangle | -\alpha \rangle | -\beta
\rangle \bigg) ,
\end{eqnarray}
with $\alpha = g_{a} t /2$ and $\beta = g_{b} t /2$. If $g_a =
g_b$, we would have $\alpha = \beta$. Equation~(\ref{twomodecat})
describes an elaborate tripartite entangled state involving one
microscopic and two mesoscopic systems. If we measure the atomic
state in the bare basis $\{ | g \rangle , | e \rangle \}$, we will
find the field in the so-called entangled coherent
states~\cite{Agarwalpuri}
\begin{eqnarray}
\label{entangledcoherent} && {\cal N}^{\pm}_{ab} \big( e^{-i
\Omega t} | \alpha e^{-i \omega t} \rangle | \beta e^{-i \omega t}
\rangle \pm e^{i \Omega t} | -\alpha e^{-i \omega t} \rangle |
-\beta e^{-i \omega t} \rangle \big) , \nonumber \\ &&
\end{eqnarray}
respectively. These states have recently been proposed as an
important tool in theory and experiments relating to the field of
quantum information~\cite{Kim}. Similar states were proposed
recently~\cite{davidovich}, entangling the vacuum and a coherent
state in two different cavities.

As before, Eq.~(\ref{tildetwomodehamilton}) is taken in the limit
where $\delta_a = \delta_b = \pm 2 \Omega$. The resulting
Hamiltonians are
\begin{eqnarray}
\label{twomodefinalhamilton2} H_{ab}^{(+)} = && \frac{\hbar g}{2}
\sum_{j=1}^N \bigg\lbrack | +_{j} \rangle \langle -_{j} | \, ( a +
b ) + | -_{j} \rangle \langle +_{j} | \, ( a^{\dagger} +
b^{\dagger} ) \bigg\rbrack , \nonumber
\\ H_{ab}^{(-)} = && \frac{\hbar g}{2}
\sum_{j=1}^N \bigg\lbrack | -_{j} \rangle \langle +_{j} | \, ( a +
b ) + | +_{j} \rangle \langle -_{j} | \, ( a^{\dagger} +
b^{\dagger} ) \bigg\rbrack \nonumber , \\
\end{eqnarray}
with $g_a = g_b$. As will be seen, these Hamiltonians are able to
produce other kinds of interesting nonclassical states. For
example, if at $t=0$, and with $N = 1$, the atom-field state is $|
+ \rangle | 0 \rangle | 0 \rangle$, then, after a time interval
$t$ of the $ H_{ab}^{(+)}$, the atom-field state will be
\begin{eqnarray}
\cos (\sqrt{2} gt) | + \rangle | 0 \rangle | 0 \rangle - i \sin
(\sqrt{2} gt) | - \rangle \frac{(| 0 \rangle | 1 \rangle + | 1
\rangle | 0 \rangle)}{\sqrt{2}} .
\end{eqnarray}
It is easy to show that if the interaction time $\tau = \sqrt{2}
\pi / 4 g$ is taken, the final state will be
\begin{eqnarray}
\frac{1}{\sqrt{2}}(| 0 \rangle | 1 \rangle + | 1 \rangle | 0
\rangle).
\end{eqnarray}
This state is a maximally entangled state between the two cavity
modes in the $\{| 0 \rangle , | 1 \rangle \}$ subspace. This
entangled state was recently produced in cavity QED with a
sequence of differently tuned interactions of a single atom with
two cavity modes~\cite{Haroche}.
Equation~(\ref{tildetwomodehamilton}) may also accept the limit
$\delta_a = - \delta_b = \pm 2 \Omega$, combining JC evolution in
the one mode and anti-JC evolution in the other one, producing
states that are beyond our present scope. Note that a number of
other interesting situations can arise by choosing nonzero
detuning between the external field and the atomic transition
frequency ($\Delta \neq 0$).

So far, all previous results are suitable for implementation in
the microwave and optical regimes in cavity QED experiments, with
atoms flying through the cavities or conveniently trapped inside
them. The microwave regime in the strong coupling limit, involving
high-$Q$ cavities and long-living Rydberg atomic levels, is more
adequate when considering the production and measurement of the
proposed nonclassical states. The optical regime, even if enjoying
more restricted strong coupling conditions, may profit from the
newly designed interactions, in particular
Eqs.~(\ref{finalhamilton1}) and (\ref{finalhamilton2}), in an
experimental setup that comfortably incorporates external
driving~\cite{Rempe}. Open cavities, in both regimes, offer the
best possibilities of experimental realization, insofar as direct
illumination of the atoms with an external classical field is
concerned. Nevertheless, in the case of closed microwave cavities,
an external field directly coupled to a second normal cavity mode
could be used. Moreover, the external field could be coupled to
the same cavity mode, profiting from a known equivalence with the
present scheme~\cite{Agarwal}. This alternative would turn the
interactions described in Eqs.~(\ref{finalhamilton2}) into a
practical realization of a closed high-$Q$ cavity plus two Ramsey
zones (without them), thus combining the advantages of these two
powerful tools. In the case of an open cavity, one can envisage
adding another microwave cavity, always transversal to the
crossing atoms, which would keep them continuously driven,
satisfying the requirements of the proposed scheme. Spatially
narrow bunches of atoms sent through microwave cavities have
already been implemented in the
laboratory~\cite{GarchingBunch,ParisBunch}.

Entanglement involving mesoscopic states is known to be very
sensitive to decoherence. Nevertheless, the interplay between a
continuously pumped entangled state, as the ones discussed in this
work, and decoherence processes might lead to a nontrivial
dynamics towards an atom-field steady-state. These considerations
are under current research and will be published elsewhere.

The question of possible implementation of these ideas in the
context of trapped ions is relevant~\cite{Daniel}. A careful study
of Eq.~(\ref{basichamilton}) shows that this interaction could be
implemented in a simple way, by means of simultaneous carrier and
red sideband excitation, as currently achieved in the laboratory.
The condition of strong driving for the carrier could be easily
satisfied and the red sideband excitation would require the
Lamb-Dicke regime.

It has been shown that an additional driving field in the usual
cavity QED experiments, and also in trapped ion systems, can be an
important tool for generating multipartite nonclassical states. We
expect that accessibility to these broader class of entangled
states, from which we presented a reduced number, offers a deeper
insight and stimulates the active field of multipartite
entanglement.

\end{document}